\documentclass[12pt,preprint]{aastex}

\usepackage{ragged2e}
\justifying\let\raggedright\justifying

\usepackage{threeparttable}

\usepackage{graphicx}
\usepackage{epsfig}
\usepackage{bm}
\usepackage{subfigure}
\usepackage{multicol}
\usepackage{multirow}
\usepackage{array}
\newcommand{\PreserveBackslash}[1]{\let\temp=\\#1\let\\=\temp}
\newcolumntype{C}[1]{>{\PreserveBackslash\centering}p{#1}}
\newcolumntype{R}[1]{>{\PreserveBackslash\raggedleft}p{#1}}
\newcolumntype{L}[1]{>{\PreserveBackslash\raggedright}p{#1}}

\shortauthors{Song et. al} \shorttitle{White-light flare associated with filament eruption}

\begin{document}

\title{Observations of a white-light flare associated with a filament eruption}

\author{Y. L. Song \altaffilmark{1,2}, Y. Guo\altaffilmark{2,3}, H. Tian\altaffilmark{1}, X. S. Zhu\altaffilmark{4}, M. Zhang\altaffilmark{5,6}, Y. J. Zhu\altaffilmark{1}}

\altaffiltext{1}{School of Earth and Space Sciences, Peking University, Beijing 100871, China; \\ Email: huitian@pku.edu.cn}
\altaffiltext{2}{Key Laboratory for Modern Astronomy and Astrophysics (Nanjing University), Ministry of Education, Nanjing 210023, China}
\altaffiltext{3}{School of Astronomy and Space Science, Nanjing University, Nanjing 210023, China}
\altaffiltext{4}{Max Planck Institute for Solar System Research, Justus-von-Liebig-Weg 3, 37077 G\"ottingen, Germany}
\altaffiltext{5}{Key Laboratory of Solar Activity, National Astronomical Observatories, Chinese Academy of
 Sciences, Beijing 100012, China}
\altaffiltext{6}{School of Astronomy and Space Science, University of Chinese Academy of Sciences, Beijing 100049, China}

\begin{abstract}
We present observations of an M5.7 white-light flare (WLF) associated with a small filament eruption in NOAA active region 11476 on 2012 May 10. During this flare, a circular flare ribbon appeared in the east and a remote brightening occurred in the northwest of the active region. Multi-wavelength data are employed to analyze the WLF, including white light (WL), ultraviolet, extreme ultraviolet, hard X-ray (HXR) and microwave. A close spatial and temporal relationship between the WL, HXR and microwave emissions is found in this WLF. However, the peak time of the WL emission lagged that of the HXR and microwave emissions by about 1-2 minutes. Such a result tends to support the back-warming mechanism for the WL emission. Interestingly, the enhanced WL emission occurred at the two footpoints of the filament. Through forced and potential field extrapolations, we find that the three-dimensional magnetic field in the flare region has a fan-spine feature and that a flux rope lies under the dome-like field structure. We describe the entire process of flare evolution into several steps, each producing the sequent brightening below the filament, the circular flare ribbons and the WL enhancement respectively. We suggest that reconnection between the magnetic field of the filament and the overlying magnetic field or reconnection within the flux rope leads to the WL enhancement.

\end{abstract}

\keywords{Sun: activity --- Sun: chromosphere --- Sun: photosphere --- Sun: flares --- Sun: magnetic fields}


\section{Introduction}

Solar flares are one of the most energetic phenomena on the sun. Flares are usually associated with a great amount of energy release and sometimes accompanied by coronal mass ejections (CMEs). Since the discovery of a solar flare on 1859 September 1 by Carrington, a large number of flares have been observed and analyzed by solar physicists. However, compared to the total number of solar flares, the number of white-light flares (WLFs) recorded in literature is very small. WLFs are usually defined as flares with an enhancement of emission in the visible continuum (Neidig et al. 1993). Up to the beginning of this century, only about 150 WLFs have been reported conclusively (Fang et al. 2013). Though WLFs are rare from observations, they are important because they challenge our knowledge on the energy transportation (Neidig 1989) and on the heating mechanisms of the lower solar atmosphere (Ding et al. 1999a). White-light emissions also provide important information on the depth and strength of the lower atmosphere heating by a flare (Kowalski et al. 2017). Several heating mechanisms have been proposed to explain WL emissions, such as electron beam bombardment  (Hudson 1972, Aboudarham \& H\'enoux 1986), Alfv\'en wave dissipation (Fletcher \& Hudson 2008), back-warming (Machado et al. 1989, Metcalf et al. 1990, Heinzel \& Kleint 2014) and chromospheric condensation (Gan et al. 1994, Kowalski et al. 2015).

The emission mechanisms and origins of WL enhancement are controversial. One suggestion is that the WL enhancement is produced by hydrogen recombination in the chromosphere (Neidig \& Wiborg 1984). This mechanism will lead to flare spectra characterized by a Balmer jump, very strong Balmer lines and WL enhancement almost synchronous with HXR or microwave emissions. Others think that the WL enhancement is produced by the increased H$^{-}$ emission in the photosphere (Hiei 1982, Boyer et al. 1985). This mechanism will lead to black body-like continuum emission, and the WL emission may be slightly delayed with respect to the HXR emission (Ding et al. 1999a, 1999b). Correspondingly, WLFs are usually classified into two types based on the spectral characteristics and the relationship between WL emission and HXR or microwave emission (Machado et al. 1986,  Fang \& Ding 1995, Ding et al. 1999a, 1999b).

In some studies, WL kernels are found to be correlated with the HXR sources spatially and temporally (Hudson et al. 1992; Metcalf et al. 2003; Chen \& Ding 2005, 2006; Krucker et al. 2011; Hao et al. 2012; Cheng et al. 2015; Kuhar et al. 2016; Yurchyshyn et al. 2017). These flares usually belong to type-I WLFs. In a relatively small number of WLFs, the WL enhancement and HXR or microwave emission do not show a strong correlation. These flares are classified as type-II WLFs. Ryan et al. (1983) reported $\gamma$-ray emission in a WLF. They found that the X-ray and $\gamma$-ray emissions peak approximately 1 minute before the WL emission, and that approximately 73\% of the WL emission occurred in a period with little or no X-ray or $\gamma$-ray emission. Ding et al. (1994) studied a WLF and found that the peak of WL emission preceded the peaks of HXR and radio bursts. Sylwester \& Sylwester (2000) studied five WLFs observed by \emph{Yohkoh} and found that WL emissions and HXR, soft X-ray emissions are not co-spatial. Recently, Hao et al.(2017) reported a circular WLF and found two types of WLF kernels: impulsive and gradual. Among the gradual kernels, one is associated with no obvious HXR source. And the maximum emission of another gradual WLF kernel occurred even 3 minutes after the peak of the HXR emission.

WLFs on the limb of the Sun provide good opportunities to investigate the height of the WL emission. Mart\'inez Oliveros et al. (2012) found that the mean heights of a HXR source and WL emission above the photosphere in a limb flare are 305 $\pm$ 170 km and 195 $\pm$ 70 km, respectively. Krucker et al. (2015) studied three WLFs at the limb and found that the centroidal heights of HXR ($\ge30$ keV) and WL sources are all about 300$-$450 km above the limb. Some studies also estimated the deposited energy by accelerated electrons and found it sufficient to produce visible continuum emission (Kleint et al. 2016, Huang et al. 2016, Lee et al 2017, Kuhar et al. 2016). However, all aforementioned WLFs are X- or M-class flares. WLFs with a lower class (e.g., C-class flares)  were also discovered with the increasing sensitivity of detectors. Using the observations of the \textit{Transition Region and Coronal Explorer} (\textit{TRACE}; Handy et al. 1999), Hudson et al. (2006) detected small enhancement of WL emission (about 8\%) in a C1.6 flare. Using the one-meter Swedish Solar Telescope (SST) at the wavelength around 3954 \AA\, Jess et al. (2008) detected strong WL emission in a C2.0 flare with a peak intensity about 300\% above the quiescent flux within a short duration and in a small region. Despite these recent progresses, many questions still remain to be answered.

Magnetic topological structure is believed to play an important role in solar eruptions (e.g. Zhang \& Low 2005, Zhang et al. 2006, 2012, Zhang \& Flyer 2008, Guo et al. 2017). It is unclear whether the magnetic field topology plays a significant role in the triggering of WLFs or not. However, it appears natural to imagine that special magnetic field configurations of a flare may store a large amount of energy, release magnetic energy in a lower height and heat deeper layers of the flaring atmosphere more easily. Therefore, if we want to obtain a complete picture of WLFs, the magnetic field structure should be investigated.

In this paper, we present multi-instrument observations of an M5.7 WLF occurred in NOAA active region (AR) 11476 on 2012 May 10. Interestingly, the WL enhancement occurred in the footpoint regions of a small filament. A three-dimensional magnetic field structure is reconstructed through forced and potential field extrapolations. Combining the obtained magnetic field configuration and multi-wavelength observations, we study the occurring process of the WLF and discuss possible mechanisms responsible for the WL enhancement. 


\section{Observations}

NOAA AR 11476 was located in the northern hemisphere. This is a very complicated AR and the configuration of the sunspot group is $\beta\gamma\delta$. During its passage on the solar disk (E$50^\circ - $W$50^\circ$), more than 50 flares were produced. However, most of them are C-class flares and only 6 are M-class flares. There is no X-class flare among them. The WLF in this study, i.e., the M5.7 flare peaked at 04:18 UT on 2012 May 10, is the biggest flare in this AR. Figure 1 presents the coronal emission, photospheric emission and magnetogram of this AR observed at 04:00 UT.

We utilize multi-wavelength data from observations of both space and ground-based telescopes for our study. The full-disk continuum filtergrams, line-of-sight (LOS) and vector magnetic field taken by the Helioseismic and Magnetic Imager (HMI; Scherrer et al. 2012) on board the \textit{Solar Dynamics Observatory} (\textit{SDO}) are used to identify the WL enhancement, study the magnetic field evolution and investigate the magnetic configuration of the flare, respectively. These data are obtained by using the spectral line of Fe {\footnotesize I} 6173 \AA\ . The temporal cadence is 45 s for the full-disk continuum filtergrams and LOS magnetograms. And it is 720 s for the vector magnetic field data. The spatial resolution of these data is about $1^{\prime\prime}$. Multi-wavelength images taken by the Atmospheric Imaging Assembly (AIA; Lemen et al. 2012) on board \textit{SDO} are used to analyze the occurring process of the flare. The cadence is 12 s for the extreme ultraviolet (EUV) passbands and 24 s for the ultraviolet (UV) passbands. And the spatial resolution is about  $1.5^{\prime\prime}$.

The Optical and Near-infrared Solar Eruption Tracer (ONSET) is a multi-wavelength solar telescope of Nanjing University, which is designed to trace solar eruptions by recording simultaneously the activities in the corona, chromosphere and photosphere (Fang et al. 2013). The corresponding wavelength windows are He {\footnotesize I} $10830 \pm 2.5$ \AA\ , H$\alpha$ ($6562.8 \pm 1.5$ \AA), and white light at 3600 \AA\ and 4250 \AA\ . Similar to HMI continuum filtergrams, observations at the wavelengths of 3600 \AA\ and 4250 \AA\ are employed to detect the WL emission. The WL data of ONSET are full-disk, with a pixel size of about $1^{\prime\prime}$ and a temporal cadence of 1 minute.

The Nobeyama Radioheliograph (NoRH; Nakajima et al. 1994) and the Nobeyama Radio Polarimeters (NoRP; Nakajima et al. 1985) at the Nobeyama Radio Observatory also observed this flare. NoRH is a radio telescope which observes the full solar disk at the frequencies of 17 GHz (intensity and circular polarization) and 34 GHz (intensity). The spatial resolution is about $10^{\prime\prime}$ for the 17 GHz images,  and $5^{\prime\prime}$ for the 34 GHz images. The temporal resolution is 0.1 s for event observations and 1 s for routine observations. NoRP observes the Sun with multiple frequencies in the microwave range, from which we can obtain the total radiation flux from the Sun. The observing frequencies are 1, 2, 3.75, 9.4, 17, 35 and 80 GHz. The observing cadence is 0.1 s.

Soft X-ray (SXR) flux (1--8 \AA) obtained by the \emph{Geostationary Orbiting Environmental Satellites} (\emph{GOES}; Grubb 1975, Neupert 2011) is employed to characterize the temporal evolution of the flare. HXR observations from the \emph{Reuven Ramaty High Energy Solar Spectroscopic Imager} (\emph{RHESSI}; Lin et al. 2002) provide important information to understand the possible mechanisms of the WL emission. The spatial resolution of \emph{REHSSI} may reach $\sim2^{\prime\prime}.3$ for full-disk observations, and the spectral resolution is 1--10 keV over a wide energy range from soft X-ray to $\gamma$-ray (Lin et al. 2002). In our study the HXR images are reconstructed from \textit{RHESSI} observations using the clean method (H\"{o}gbom 1974; Hurford et al. 2002) in the energy bands of 12--25 keV, 25--50 keV and 50--100 keV. HXR full-disk fluxes of different energy bands are also used to investigate the temporal relationship with the WL emission.

\section{The relationship between WL, HXR and microwave emissions}

The spatial relationship between the WL, HXR and microwave emissions is shown in Figure 2. The top three panels in Figure 2 are the AIA 94 \AA\ , 171\AA\ and 1700 \AA\  images of NOAA AR 11476 at 04:16 UT, close to the peak time of the flare. Contours of HXR and microwave emissions are overlaid on them. From these panels we find two regions of interest. One region is marked by the white box and shows significantly enhanced emission in the UV and EUV passbands. The flare mainly occurred in this region. The other region is located to the northwest of the flaring region, showing a remote brightening. The middle panels in Figure 2 present the photosphere filtergrams observed by ONSET 3600 \AA\ , 4250 \AA\ and HMI 6173 \AA\ at 04:15 UT before the WL enhancement occurred.  Regions R2 and R1 marked by the white boxes in Figure 2(f) refer to the region of WL enhancement and a quiet region, which are selected to calculate the WL emission flux and estimate the uncertainty, respectively. The bottom panels are the WL difference images showing the WL enhancement during the flare. It should be noted that the difference images are obtained through the calculation of $(I-I_0)/I_0$,  where $I$ refers to the WL intensity at the peak time of WL emission and $I_0$ refers to the WL intensity before WL enhancement. The enhancement of WL emission can be seen in all the three WL channels. The enhancement in HMI 6173 \AA\ is slightly stronger than that in ONSET 3600 \AA\ and 4250 \AA\ . Such a difference may be caused by the fact that these different instruments and filters sample emission at different heights of the solar atmosphere and that their spatial resolutions are different. Contours of HXR and microwave emissions are overlaid on the difference images, from which we can see that the WL enhancement appeared nearly in the center of the HXR and microwave sources. In other words, the WL enhancement spatially coincided with the HXR and microwave emissions very well.

Figure 3 shows the temporal variations of the microwave, white-light, Soft and Hard X-ray fluxes. Because the cloud affected the observations of ONSET during the flare, we only use the HMI continuum data to study the temporal evolution of the WL emission. To estimate the error in the calculation of the WL enhancement, we obtain a WL curve by averaging the intensity in the quiet region R1 ($10^{\prime\prime}\times10^{\prime\prime}$, see Figure 2(f)) during the flare. Then we calculate the standard deviation of the curve, which is used as the measurement error of WL emission in the flaring region R2 (see Figure 2(f)). Following several previous studies (e.g., Hudson et al. 2006, Kuhar et al. 2016), the relative enhancement of the WL emission is defined as the maximum value of $(I-I_0)/I_0$, where $I_0$ and $I$ refer to the HMI continuum intensities at 04:15:50 UT (before the WL enhancement) and 04:17:20 UT (around the peak of the WL emission), respectively. The relative enhancement turns out to be $0.26\pm 0.02$ in this flare. Figure 3 shows that the WL emission peaked slightly after the HXR and microwave peak time. The peak time for the HXR and microwave emission is around 04:16:30 UT, while for the WL emission it is close to 04:18:00 UT. Considering the 45 s cadence of the HMI continuum images, we conclude that the peak time of the WL emission lags the peak time of microwave and HXR fluxes by about 1--2 minutes. This time lag appears to support the back-warming mechanism for the WL enhancement, i.e., non-thermal electrons deposit energy in the chromosphere, and the WL emission is produced through back-warming from the chromosphere to the photosphere. A time lag may be expected between the arrival of the non-thermal electrons in the chromosphere and the production of sufficient amount of hot plasma. Electrons with much higher energy might directly penetrate to the even lower atmosphere and produce the WL emission. However, in this scenario there should be no significant time lag between the WL and HXR emissions (Rust \& Hegwer 1975, Neidig \& Zirin 1985). The Alfv\'en wave dissipation mechanism may produce a similar time lag. However, we do not find any evidence to support the existence of  Alfv\'en waves in our flare. It is also noteworthy that the WL emission peaked at the same time of the SXR peak in this flare.

Figure 4 shows the \emph{RHESSI} energy spectrum acquired during the time interval of 04:16:20 -- 04:16:40 UT, close to the peak of the flare. The spectrum is fitted by a variable thermal function (green) and a non-thermal thick-target bremsstrahlung function (yellow). The emission measure (EM) and temperature (T) for the thermal component are $\sim2.68\times10^{48}$ cm$^{-3}$ and $\sim 25.4$ MK, respectively. The spectral index ($\delta$) for the HXR electrons is $\sim3.25$. To calculate the total energy flux of the non-thermal electrons above a cutoff energy ($E_c$) under the thick-target approximation (Hudson et al. 1978; Watanabe et al. 2010a), we use the IDL routine of \emph{calc\_nontherm\_electron\_energy\_flux.pro} in \emph{SSW}. And the correspongding input parameters (total integrated electron flux; low delta, i.e., index of electron distribution function below break;  break energy; high delta, i.e., index of electron distribution function above break; low energy cutoff; high energy cutoff) for this method can be seen in the spectra fitting results shown in Figure 4. And we determine the area of the HXR emission region as pixels where the integrated HXR intensity ($50-100$keV) is greater than 60\% of the maximum intensity. In our study the HXR emission area is $\sim1.19\times10^{18}$cm$^2$, and the nonthermal electron energy flux is $\sim 4.76 \times 10^{10}$ erg $\cdot$ cm$^{-2} \cdot $s$^{-1}$.

\section{WL emission and magnetic field configuration}

Figure 5 shows the location of WL enhancement in HMI and AIA images. We first calculate the standard deviation of the running difference in the quiet-sun region R1 during the flare and the average value is about 0.013, which represents a level of the fluctuation of the background. Then we identify the WL enhancement regions as the areas where the emission change ($(I-I_0)/I_0$) is greater than 0.05 ($\sim4\times0.013$). From Figure 5(a), we find mainly two regions of WL enhancement, one is mainly inside the umbra of the negative-polarity sunspot and the other is located mainly in the penumbra of the positive-polarity sunspot. Figure 5(b) shows a filament rooted in the two sunspots. The WL enhancement in the umbra of the negative sunspot is undoubtedly located at the footpoint region of the filament. The other patch of WL enhancement is not strictly located at the footpoint of the filament, although it is very close. As shown in Figure 5(c), the transverse magnetic field in the negative-polarity sunspot possesses a strong rotating pattern, indicating a strong twist of the filament rooted there. Note that one patch of WL enhancement is just located in the region where the transverse magnetic field shows the strongest rotating pattern. This implies that there may be a close relationship between the WL enhancement and the strongly twisted filament. The horizontal velocity field in Figure 5(d) is obtained by the Differential Affine Velocity Estimator (DAVE; Schuck 2006). It shows an apparent rotating motion in the north part of the negative-polarity sunspot, which may explain the observational fact that several filaments formed and erupted in this region during a few days. It is known that the rotation of sunspots plays an important role in the formation of some filaments (Yan et al. 2015, 2016).

To study the magnetic field structures associated with the WLF, we reconstruct the 3D magnetic field through forced and potential field extrapolations (Figure 6). Figure 5 shows that the filament is small and that its length is less than $30^{\prime\prime}$. We apply a forced-field extrapolation (FFE; Zhu et al. 2013, 2016) to reconstruct the magnetic field structure around the filament. This method computes the magnetic field by solving full MHD equations using the ``stress-and-relax" approach. It is particularly appropriate for the magnetic field extrapolation in the lower solar atmosphere, such as the chromosphere, transition region and low corona where the plasma $\beta$ is relatively high (Zhu et al. 2013, 2016). To reconstruct the higher background magnetic field structures, we use the potential field extrapolation. Both reconstructions are based on the vector magnetic fields observed by HMI. In this study, the extrapolation using the FFE method is performed in a box of $864\times576\times161$ grid points with a spatial resolution of $0.5^{\prime\prime}$ per grid, and the extrapolation through the potential field model is performed in a box of  $244\times196\times196$ grid points with a resolution of $2^{\prime\prime}$ per grid. As shown in Figure 6(a), the extrapolated flux rope is rooted in the two main sunspots, which coincides exactly with the observed filament. Figure 6(b) shows that there is a dome-like magnetic field structure above the flux rope. Figure 6(c) presents the magnetic field configuration in a lager field of view. It shows a typical fan-spine structure (e.g. Sun et al. 2013, Mandrini et al. 2014, Yang et al. 2015), where a flux rope is located below the dome. And we find a null point at a position of (39.7, 54.3, 3.87) in the grid-point box of the potential field extrapolation, located at a height of about 5.6 Mm above the photosphere. The footpoint of the inner spine and one footpoint of the flux rope are rooted in the same negative-polarity sunspot. To show this fan-spine structure more clearly, we calculate the squashing factor \emph{Q} on three different surfaces. Figure 6(d) shows the locations of these three surfaces: $x-z$ and $y-z$ planes passing through the center of the negative-polarity sunspot, and a horizontal plane at the height of $z \sim 3$ Mm.

Quasi-Separatrix Layers (QSLs) refer to thin irregular layers where the connectivity of 3D magnetic field lines changes drastically (e.g., D\'emoulin  et al. 1997, D\'emoulin  2006, Guo et al. 2013). QSLs are thought to be a suitable environment where magnetic reconnection may occur (e.g., Mandrini et al. 1997, Titov et al. 2002, Wilmot-Smith et al. 2010, Janvier et al. 2013). Figure 7 shows the QSLs on the three slices marked by the white arrows and box in Figure 6(d).  QSLs here are obtained using a 3D computation method for the squashing degree (Pariat \& Demoulin 2012; Yang et al. 2015) based on the potential field extrapolation. From the QSLs on the $x-z$ and $y-z$ planes (Figures 7(a) and 7(b)), we see a dome-like QSL. And from the Figures 7(a) and 7(b), we also see a spine-like QSL above the dome-like QSL. From the QSL on the horizontal layer ($x-y$ plane, Figure 7(c)) at a height of $z\sim3$ Mm above the photosphere, a circular QSL can be seen in the flare region. Figure 7(d) shows an AIA 1600 \AA\ image taken at 04:18 UT, the peak time of the flare. QSLs in Figure 7(c) with a level of $Q\geq10^3$ are overlaid on the 1600 \AA\ image. We find that the circular flare ribbon coincides with the circular QSL very well and that the remote brightening is located at the outer QSL. All these features suggest a standard fan-spine configuration of the magnetic field structures. We also find that the outer QSL where the remote brightening occurred is an extended large-scale QSL. Previous studies have shown that a large-scale QSL could be induced by a larger quadropolar-like magnetic fields (e.g.,D\'emoulin  et al. 1996, Aulanier et al. 2005, Yang et al. 2015), which may also be the case in our observation.

\section{The process of flare eruption}

Figure 8 presents the detailed evolution of the flare as seen in the AIA  304 \AA\ , 131 \AA\ , 94 \AA\ and 1600 \AA\ passbands. A small filament was clearly present before the flare (e.g., 04:00 UT), as shown in Figures 8(a1), (b1) and (c1). At the beginning of the flare (e.g., 04:14 UT), a brightening appeared below the filament and the filament was obviously uplifted. As the filament started to rise, reconnection between highly sheared or twisted field lines below the filament likely resulted in the brightening. Later, a vertical current sheet might form below the erupting flux rope that possibly contains filament materials. Reconnection occurring in the current sheet might also contribute to the brightening below the erupting filament in our observation. As the filament was uplifted to a certain hight (e.g., 04:15 UT), the reconnection in the null point of the fan-spine structure was triggered. As a result, a circular flare ribbon (CFR) and remote brightening appeared (Figures 8(a3), (b3), (c3) and (d3)). At this time, the flux rope can still be seen in AIA 131 \AA\ and 94 \AA\ channels from Figures 8(b3) and 8(c3). At 04:16 UT, the EUV emission became stronger and the WL enhancement began to appear. The green contours in Figures 8(a4), (b4), (c4) and (d4) mark the regions where the WL enhancement occurred. The AIA 94 \AA\ image at 04:16 UT shows that the filament still existed there but was uplifted to an even higher layer. And patches of the WL enhancement were located at the footpoints of the filament (Figure 8(c4)). After the flare (e.g., 04:40 UT), the filament vanished, likely due to reconnection with the higher background magnetic field.

Figure 9 shows the relationship between the WL enhancement and the filament eruption. Figures 9(b1)--(b5) are difference images of the HMI continuum intensity in the period of 04:15:05 UT -- 04:18:50 UT. The WL enhancement first appeared at the time around 04:16 UT and reached a peak around 04:18 UT, after which the emission began to decrease. Figures 9(a1)--(a10) are the AIA 335 \AA\ images during the same period. These images have been processed using the method of Multi-Scale Gaussian Normalization (MGN; Morgan \& Druckm\"{u}ller 2014). We see that the filament kept rising before 04:16:29 UT, and the emission of AIA 335 \AA\ became stronger and stronger around the footpoints of the two legs, which implies a process of energy release. At 04:16:53 UT the magnetic field of the filament likely reconnected with the background magnetic field, and the 335 \AA\ emission around the two footpoints was still very strong. It is worth noting that the rapid enhancement of the 335 \AA\ emission appeared when the filament reached the highest height and the magnetic field of the filament possibly began to reconnect with the background field (Figures 9(a3)--(a6)). From Figures 9(a3)--(a6), (b2)--(b3) and (c2)--(c3), it is also obvious that the WL enhancement occurred at the two footpoints of the filament. We overlay the contours of HXR fluxes from \emph{REHSSI} with the energy bands of 12--25 keV (magenta), 25--50 keV (green) and 50--100 keV (yellow) in these images. Before 04:16 UT, the HXR sources in these three channels coincided with each other well (Figures 9(a1) and (a2)). But during the period of 04:16 UT -- 04:18 UT, the 25--50 keV and 50--100 keV sources were mainly located at the low height of the eastern leg of the filament, and the 12--25 keV source was found mainly at the low height of the western leg of the filament. Based on these observations, we interpret the WL emission enhancement as being produced by energy deposition through non-thermal electrons propagating downward along the two legs of the filament, due to reconnection between the magnetic field of the filament (flux rope) and the higher background magnetic field or reconnection within the filament. We notice that the area of WL enhancement is smaller than the size of the HXR sources during the flare. The different sizes of the HXR sources and WLF kernels tend to suggest that they are produced at different layers in the atmosphere. In the scenario of backwarming, the HXR sources are located in the chromosphere where a large amount of energy is deposited. The chromospheric heating and the resultant EUV and SXR emissions may be highly nonuniform, and thus the efficiency of backwarming is likely different at different locations below the heated atmosphere. In that case, smaller and patchy WLF kernels may be expected. More observations are needed to examine whether this phenomenon is typical in WLFs or not.

A cartoon in Figure 10 summarizes the whole process of the flare. First, a flux rope lay under a dome-like magnetic field structure and it might become unstable due to plasma flows around or below it. At a certain time it lost force balance and began to rise. As time evolved, the filament was uplifted to different heights, then a brightening below the filament, circular flare ribbon and WL enhancement appeared in sequence. Finally the filament reconnected with higher background magnetic field and erupted. As we have seen in Figure 5, the transverse magnetic field in the negative-polarity sunspot possesses a strong rotating pattern, indicating a strong twist of the filament. Such a filament can easily become unstable and rise to higher layers. This interaction, or reconnection between highly sheared or twisted field lines and within the current sheet in the stretched field lines behind the rising filament, gave rise to the small-scale brightening below the filament. The rising filament pushed the inner fan magnetic field above it to the null point, where reconnection was triggered. As a result, a circular ribbon was formed and the remote brightening appeared. Consequently, the filament (flux rope) had a close contact with the overlying coronal field lines. Reconnection between them or within the legs of the filament produced high-energy electrons that eventually led to the enhanced WL emission at the two footpoints of the filament.


\section{Summary and discussion}

We present observations of an M5.7 WLF occurred in NOAA AR 11476 on 2012 May 10. Multi-wavelength observations from both space and ground-based instruments have been used to analyze this WLF. We have investigated the relationship between the WL, HXR and microwave emissions, and reconstructed the 3D magnetic field configuration of the flare. We have studied the relationship between magnetic field evolution and flare \textbf{eruption}, and investigated various stages of the flare process. We have also discussed how the WL enhancement was probably produced during the flare.

The WL enhancement spatially coincided with the HXR and microwave emission very well. The variation of WL emission also temporally correlated with the HXR and microwave fluxes. However, the peak time of the WL emission lagged the peak of the HXR and microwaves fluxes by 1--2 minutes. The relationship between the WL emission and the HXR (or microwave) if often used to classify a WLF (Machado et al. 1986, Fang \& Ding 1995, Ding et al. 1999a, 1999b). For type-I WLFs, there is a strong correlation between them. However, for type-II WLFs there is no obvious correlation. Based on this definition, the WLF we have studied \textbf{ may belong} to type-I WLF. The time lag of the WL emission with respect to the HXR/microwaves enhancement seems to support the back-warming mechanism for the WL emission. We find that the WL enhancement in this flare occurred at the footpoints of a small filament and that the energy contributing to the WL enhancement likely came from the internal reconnection of the flux rope or from the reconnection between the flux rope and the overlying magnetic field.


The magnetic field configuration of the flare is analyzed through forced and potential field extrapolations using the vector magnetic field observed by HMI. A flux rope lay along the magnetic polarity inversion line (PIL) and was rooted in two main sunspots in the flare region. Before the flare, the transverse component of the photospheric magnetic field in the negative-polarity sunspot presented a strong rotating pattern, suggesting a strong twist of the filament-associated flux rope. The magnetic field configuration of the flare shows a standard fan-spine structure and the flux rope lay just below the dome-like magnetic field. And during the flare, a circular flare ribbon and a remote brightening were clearly observed, which is also consistent with this magnetic field structure. When the reconnection in the null point occurred, the constraint from the upper background magnetic field was reduced. Under such a circumstance, the flux rope eruption and null point reconnection formed a positive feedback to each other, so the flux rope erupted faster and higher.

Based on the reconstructed magnetic field configuration and AIA observations, we find that the process of the flare can be divided into several stages (Figures 8 and 10). First, the filament eruption was triggered by some kind of instability, and a brightening appeared below the filament. Second, the circular flare ribbon and remote brightening appeared as the filament was uplifted to a higher layer. Third, the WL enhancement began to appear at the two footpoints of the filament. Finally the filament possibly reconnected with the higher background magnetic field. From AIA observations (Figure 8), the corresponding times for the second and third steps are 04:15 UT and 04:16 UT, respectively. Basically these different stages corresponded to different heights where the filament rose to, which also indicated magnetic reconnections at different locations.

It is worth noting that the WL enhancement in this study occurred at the two footpoints of the filament. HXR sources that are located in the footpoint regions have been reported before (eg. Liu \& Alexander 2009, Guo et al. 2012, Yang et al. 2015 ). However, enhanced WL emission at the footpoint areas during a flare has been rarely reported. In our observation, the WL enhancement mainly occurred between 04:16 UT and 04:18 UT. As shown in Figure 9, when the WL emission started to increase at 04:16 UT, the filament was still there but located at a relatively higher height. At 04:16:53 UT the magnetic field of the filament possibly reconnected with the background magnetic field, and the emission around the two footpoints of the filament was still very strong. The HXR emission source was mainly located below the filament and cospatial with the small-scale brightening before 04:16 UT. From 04:16 UT to 04:18 UT, the HXR sources were mainly concentrated on the lower parts of the two legs. It has been suggested that reconnection within the flux rope could happen, if the twist is large enough to result in an internal kink instability (eg. Galsgaard \& Nordlund 1997, Haynes \& Arber 2007). Besides, magnetic reconnection can also occur at the border of the flux rope. Both mechanisms may accelerate electrons and back-warm the lower atmosphere, which would result in the WL enhancement at the two footpoints of the filament. Based on these observations, we interpret the WL emission enhancement as being produced by the energy deposition through non-thermal electrons accelerated in the two legs of the filament, due to reconnection with the higher background magnetic field near the border of the filament (flux rope) or internal reconnection within the filament.

\acknowledgements
We thank Profs. M.-D. Ding, H.-M. Wang and Y.-N. Su for very helpful discussion. This work is supported by NSFC grants 11790304 (11790300), 11125314, 11773016, 11533005 and U1531247, the Recruitment Program of Global Experts of China,  the Key Laboratory for Modern Astronomy and Astrophysics (Nanjing University) and the Max Planck Partner Group program. H.T. thanks ISSI and ISSI-BJ for the support to the team ``Diagnosing heating mechanisms in solar flares through spectroscopic observations of flare ribbons''. 

\newpage

\begin{figure}[!ht]
\centerline{\includegraphics[width=0.9\textwidth]{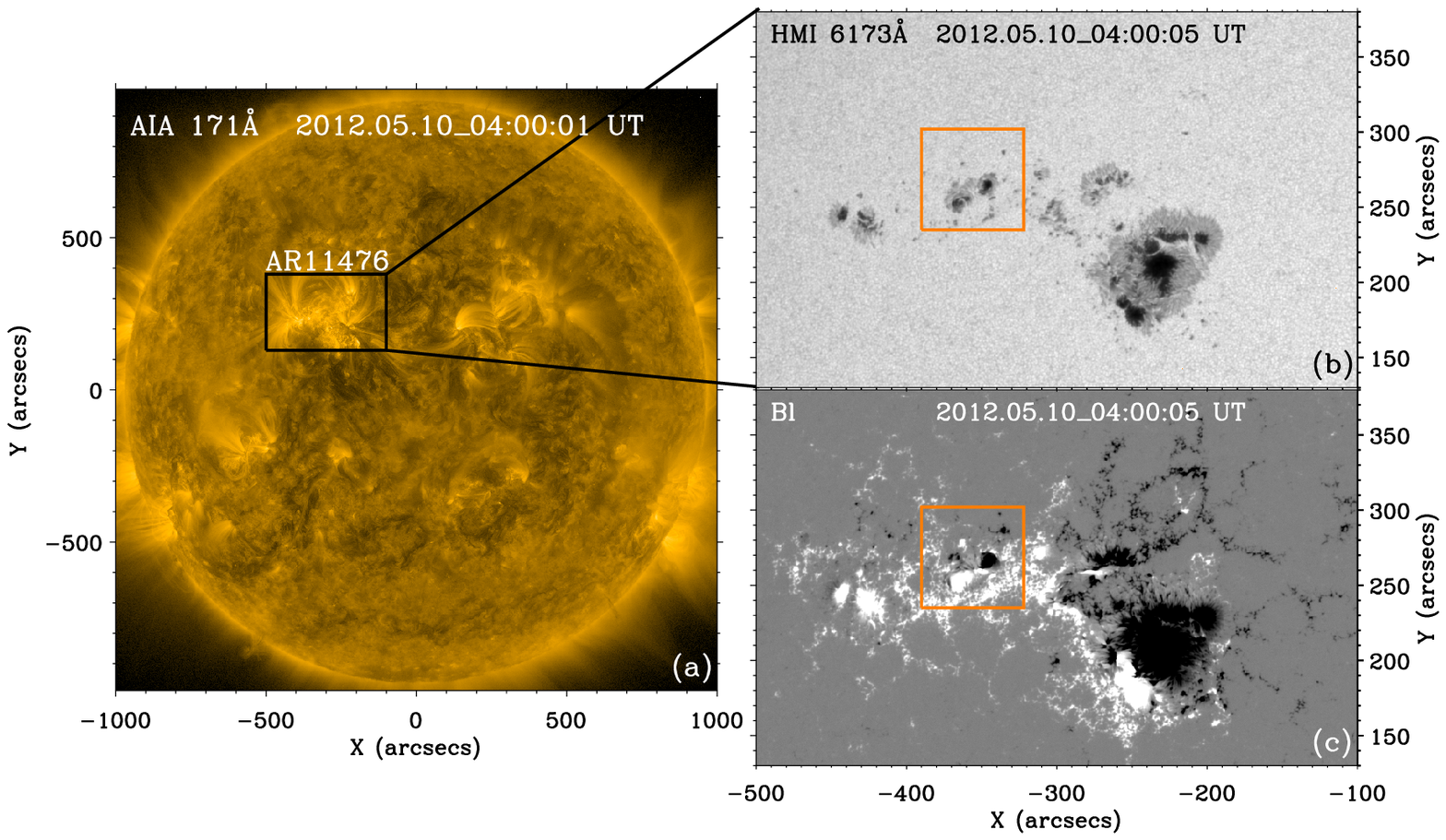}}
\caption{NOAA AR 11476. (a) AIA 171 \AA\ full-disk image taken at 04:00 UT on 2012 May 10. The black box marks AR NOAA 11476. (b) HMI 6173 \AA\ continuum intensity image of the AR taken at the same time. (c) HMI line-of-sight magnetogram ($B_l$) of the AR. Brown boxes in panels (b) and (c) mark the area where the flare occurred. }
\end{figure}

\begin{figure}[!ht]
\centerline{\includegraphics[width=0.9\textwidth]{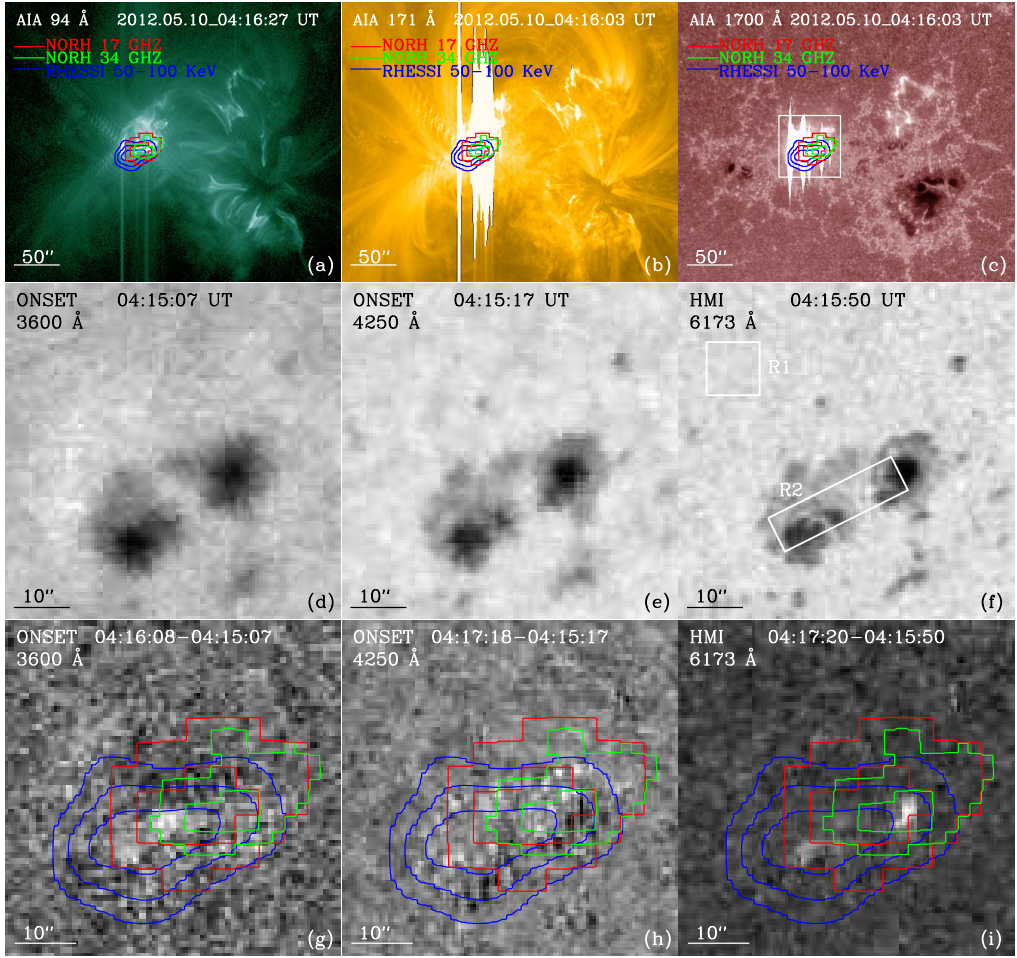}}
\caption{Spatial relationship between the white-light emission and the hard X-ray and microwave emissions. Panels (a)--(c) are the AIA 94 \AA\ , 171 \AA\ and 1700 \AA\ images taken at 04:16 UT, close to the flare peak. The white box in panel (c) corresponds to the field of view (FOV) of panels (d)--(i). Panels (d)--(f) are the ONSET 3600 \AA\ , 4250 \AA\ and HMI 6173 \AA\ images taken at 04:15 UT. Panels (g)--(i) are the difference images of these three passbands. Red and green contours in these panels represent the microwave 17 GHz and 34 GHz emissions observed by NoRH with levels of 30\% and 70\% of the maximum intensity. Blue contours represent the RHESSI 50-100 KeV emission with levels of 50\%, 70\%, and 90\% of the maximum intensity. Regions R2 and R1 in panel (f) refer to the area where we obtain the variation of the white-light emission and a quiet-sun region where we use for the estimation of the error for WL emission, respectively. }
\end{figure}

\begin{figure}[!ht]
\centerline{\includegraphics[width=0.8\textwidth]{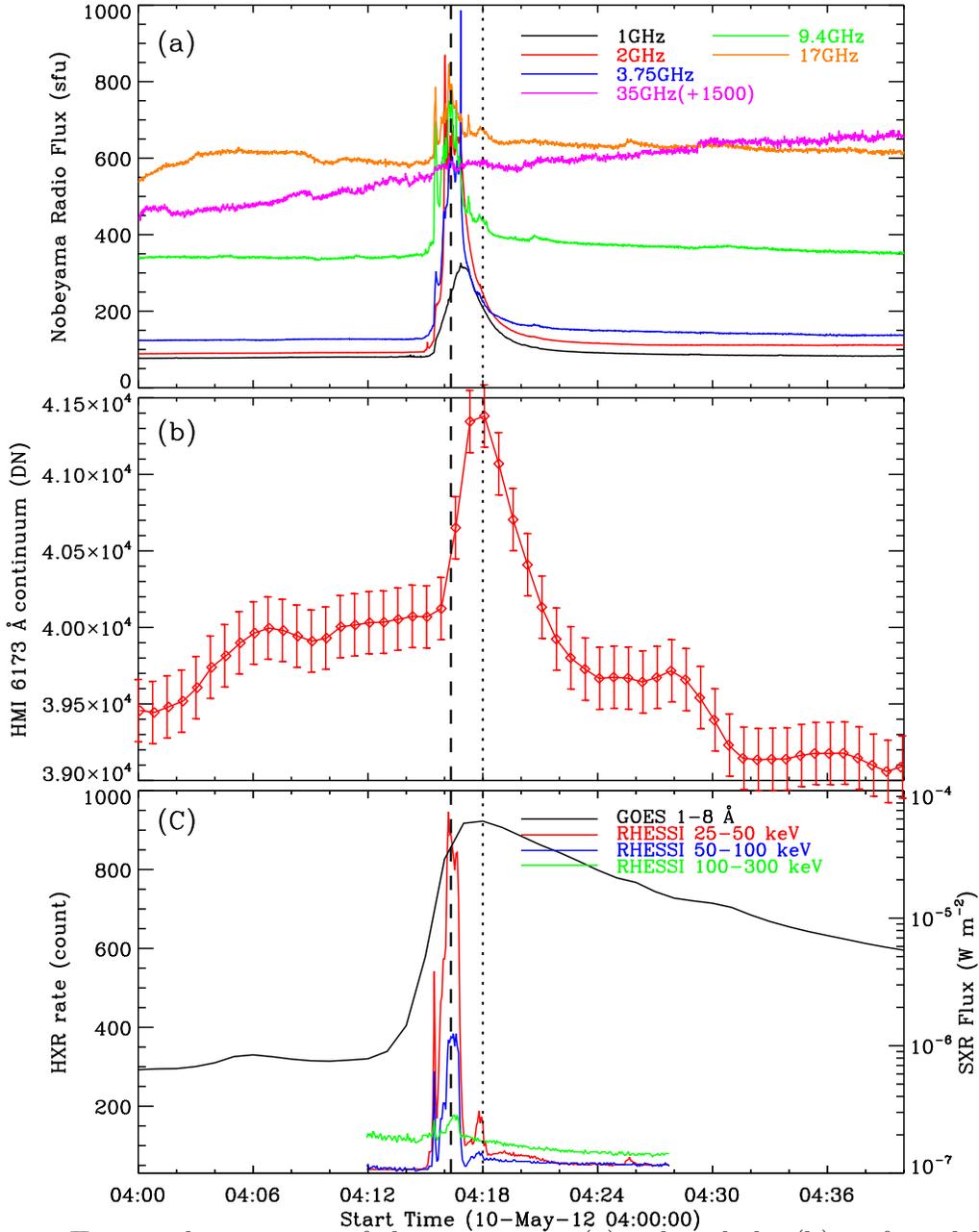}}
\caption{ Temporal variations of the microwave (a), white light (b), soft and hard X-ray (c) fluxes. The dashed line marks the peak times of microwave and hard X-ray fluxes. The dotted line marks the peak time of the white-light emission. In panel (b), the temporal variation of the HMI 6173 \AA\ continuum intensity measures the average intensity in the region R2 that is defined in Figure 2(f) .}
\end{figure}

\begin{figure}[!ht]
\centerline{\includegraphics[width=0.6\textwidth]{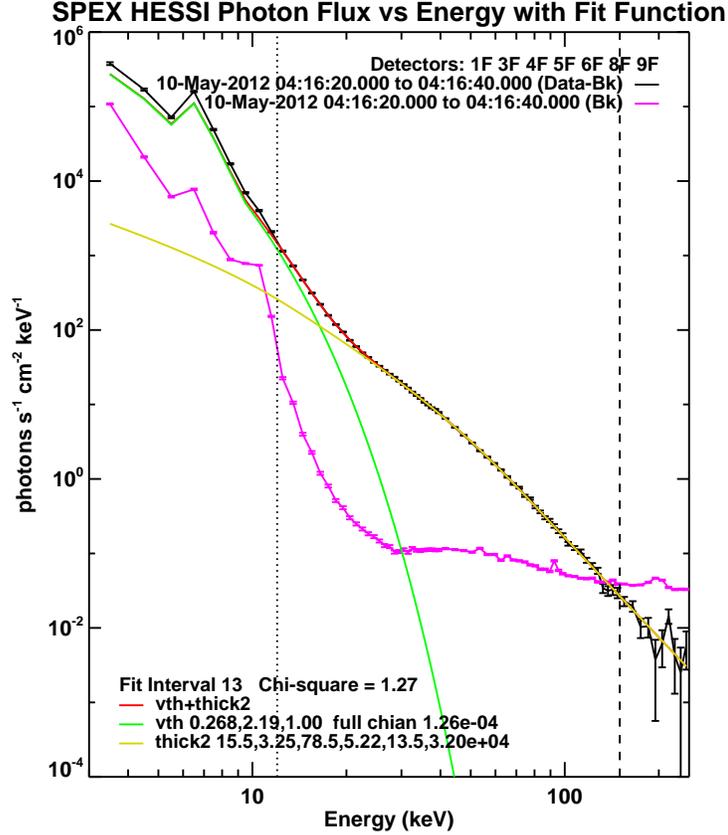}}
\caption{\emph{RHESSI} energy spectrum acquired in the time interval of 04:16:20--04:16:40 UT, around the peak time of the flare. The black curve is the observational spectral data after subtracting the background (purple curve), which is fitted by a variable thermal function (green) and a non-thermal thick-target bremsstrahlung function (yellow). The two vertical lines (dotted and dashed) indicate the fitting energy ranges. From the fitting of the spectrum, we have obtained an EM of $2.68\times 10^{48} \ $cm$^{-3}$, a temperature of 25.4 MK and a spectral index of 3.25.}
\end{figure}




\begin{figure}[!ht]
\centerline{\includegraphics[width=0.9\textwidth]{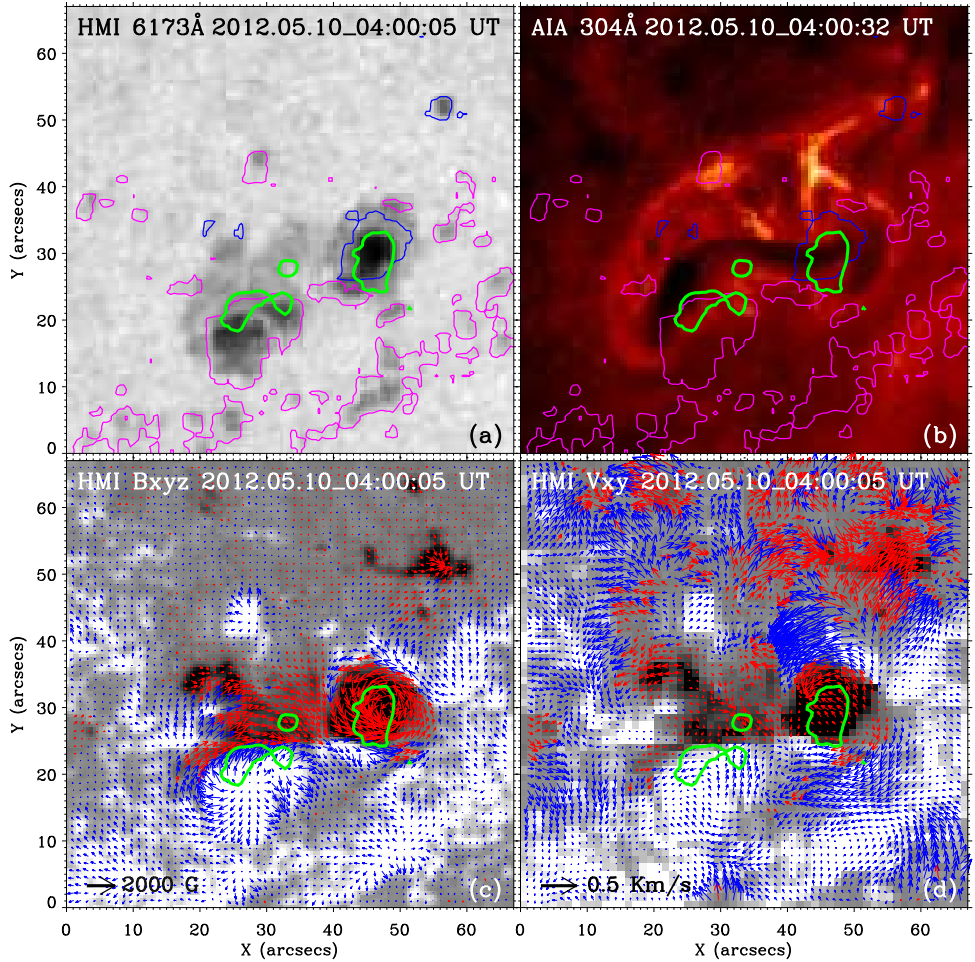}}
\caption{ Upper panels: HMI continuum intensity (a) and AIA 304 \AA\ (b) images taken at 04:00 UT on 2012 May 10. Magenta and Blue contours represent the positive and negative magnetic fields with a level of $|B_l|=500$ G.  Lower panels: HMI vector magnetic field (c) and horizontal velocity field (d) overlaid on the simultaneously taken longitudinal magnetogram. The velocity field is obtained using the method of the differential affine velocity estimator (DAVE). Red and blue arrows represent velocity vectors at locations of negative and positive fluxes, respectively. Green contours mark the regions of WL enhancement. Here the contour level is $(I-I_0)/I_0=0.05$, where $I_0$ and $I$ refer to the HMI continuum intensities at 04:15:50 UT and 04:17:20 UT, respectively.}
\end{figure}

\begin{figure}[!ht]
\centerline{\includegraphics[width=0.9\textwidth]{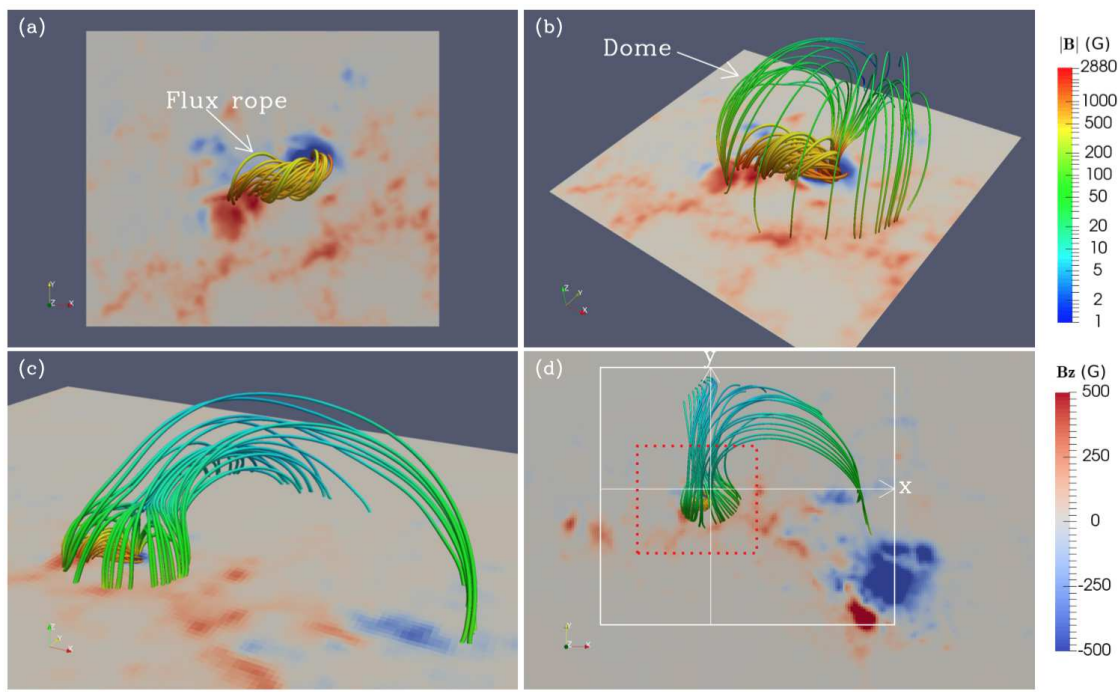}}
\caption{ Results of magnetic field extrapolations using the HMI full-disk vector magnetic field observed at 04:00 UT on 2012 May 10. (a) The flux rope obtained through the forced field extrapolation. (b) The flux rope and the dome structure above the flux rope. (c, d) The magnetic field structure from different perspectives in a larger field of view than that of (a) and (b). The magnetic fields above the flux rope are obtained through the potential field extrapolation. The red dotted box in panel (d) corresponds to the FOV of panels (a) and (b). The white box and two arrows in panel (d) mark the slices where we calculate the Quasi-Separatrix Layers (QSLs) shown in Figure 7.}
\end{figure}

\begin{figure}[!ht]
\centerline{\includegraphics[width=0.95\textwidth]{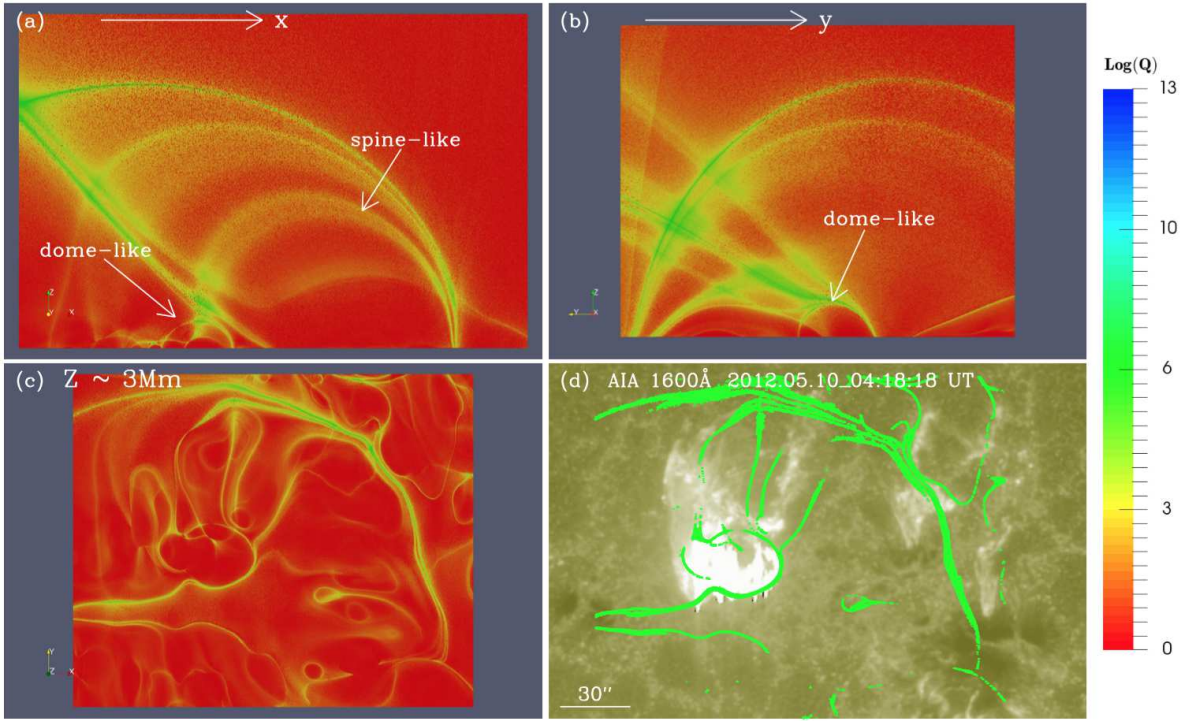}}
\caption{ QSLs on the slices shown in Figure 6(d). They are calculated based on the potential field extrapolation using the HMI full-disk vector magnetogram at 04:00 UT. (a, b) QSLs on the $x-z$ and $y-z$ planes passing through the center of the negative-polarity sunspot, respectively. A dome like structure can be seen. (c) QSL on the plane marked by the white box in Figure 6(d) at the height of $Z\sim3$ Mm. (d) AIA 1600 \AA\ image taken at 04:18 UT. The green contours correspond to the QSLs in panel (c) with a level of $Q = 10^3$.}
\end{figure}

\begin{figure}[!ht]
\centerline{\includegraphics[width=1.1\textwidth]{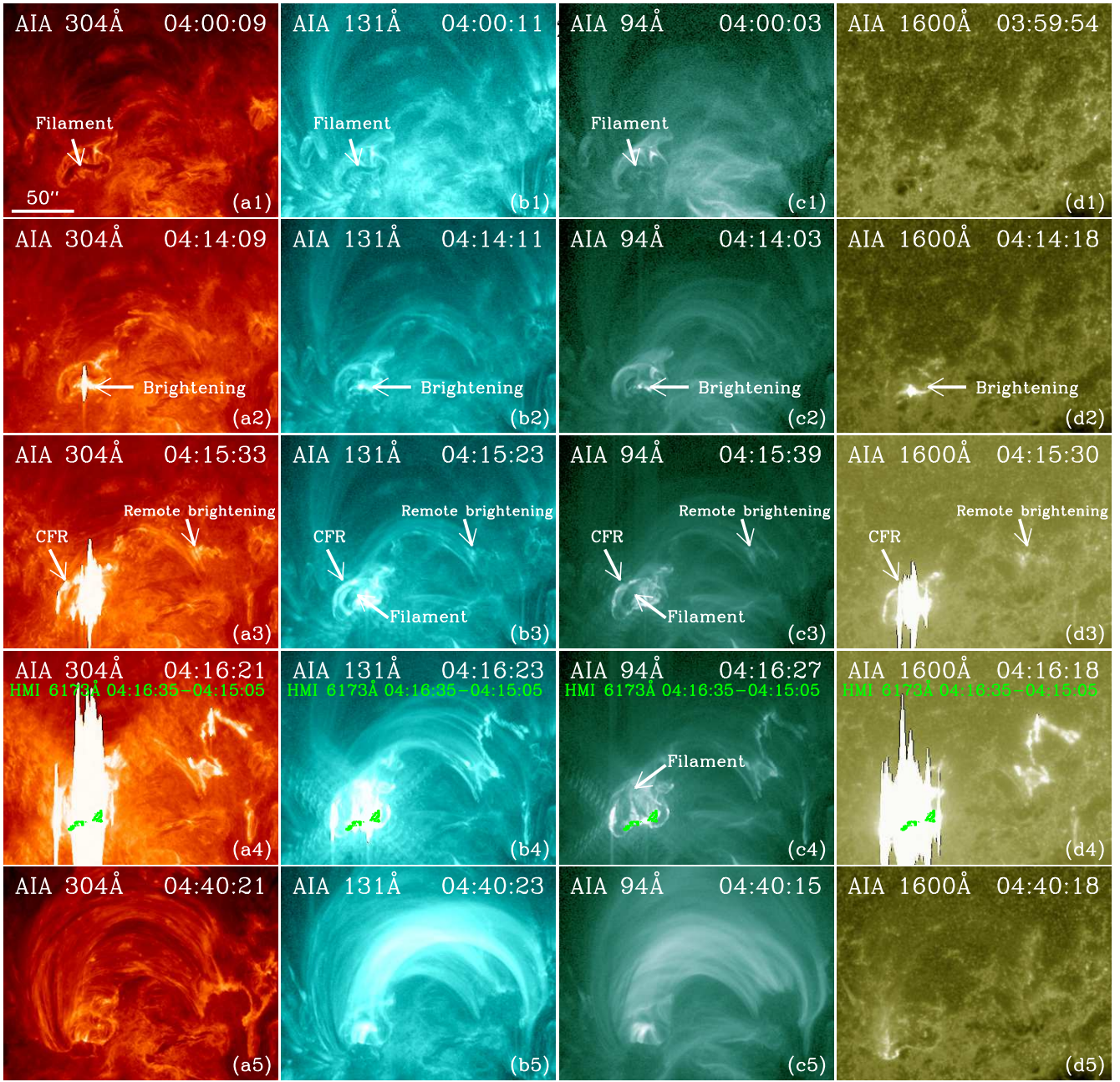}}
\caption{ Different stages of the flare. Panels (a1)--(a5), (b1)--(b5), (c1)--(c5) and (d1)--(d5) show the AIA 304 \AA\ , 131 \AA\ , 94 \AA\ and 1600 \AA\ images, respectively. For each passband we show images taken at five different times, 04:00 UT (before the flare), 04:14 UT (occurrence of a brightening below the filament), 04:15 UT (circular flare ribbon (CFR) and remote brightening appeared), 04:16 UT (WL emission appeared) and 04:40 UT (after the flare). The green patches in panels (a4)-(d4) represent locations of WL enhancement.}
\end{figure}

\begin{figure}[!ht]
\centerline{\includegraphics[width=0.9\textwidth]{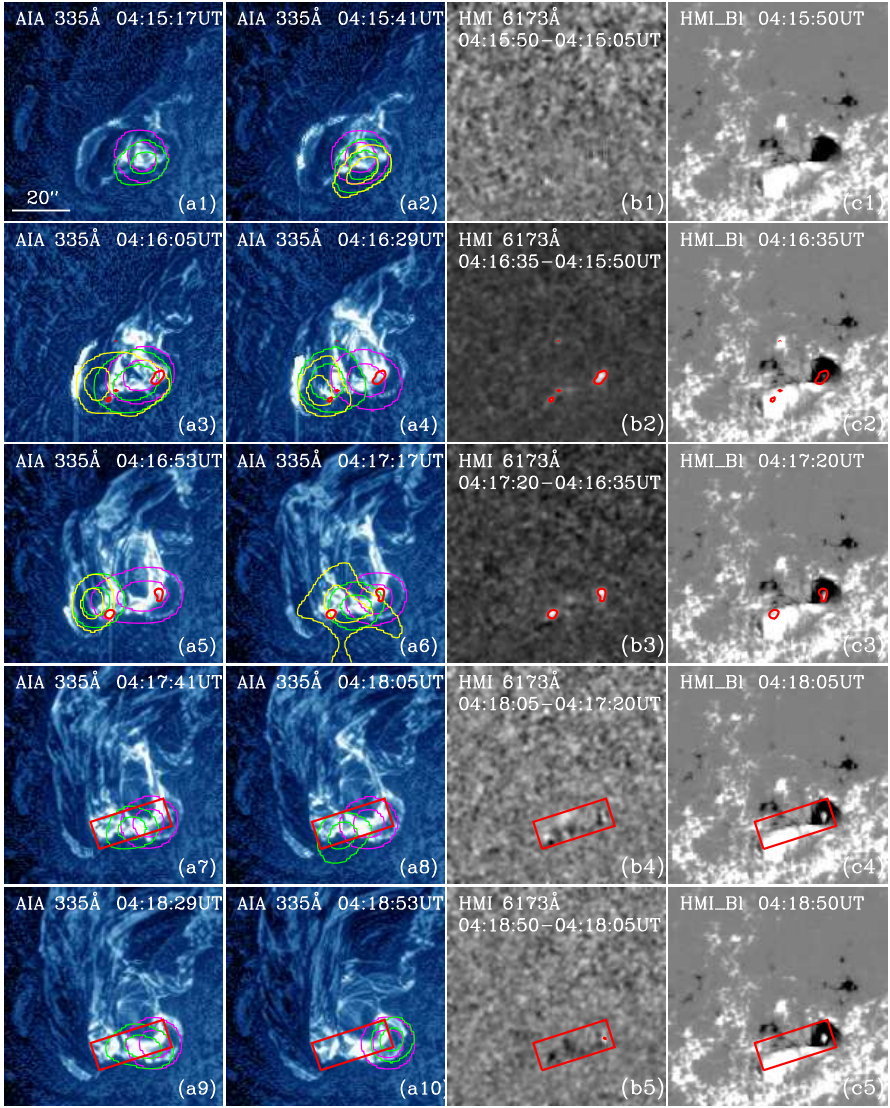}}
\caption{ (a1)-(a10) AIA 335 \AA\ images processed using the method of Multi-Scale Gaussian Normalization (MGN). (b1)-(b5) The running difference images of HMI 6173 \AA\ continuum intensity. (c1)-(c5) The HMI LOS magnetograms. All these images have the same field of view. The red box marks the area where white-light emission enhancement occurred. The magenta, green and yellow contours with levels of 70\% and 90\% of the maximum value show the \emph{REHSSI} 12--25 keV, 25--50 keV and 50--100 keV sources, respectively.}
\end{figure}

\begin{figure}[!ht]
\centerline{\includegraphics[width=\textwidth]{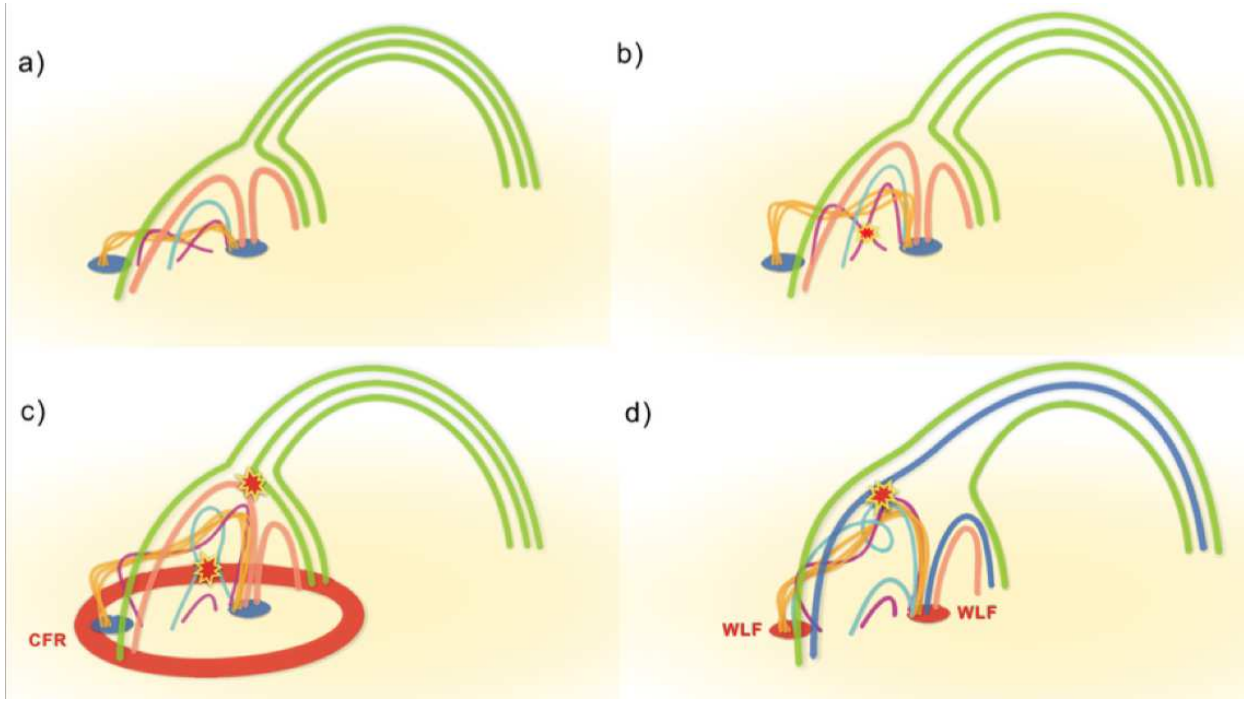}}
\caption{ A cartoon showing different stages of the flare. (a) Before the flare, there is a filament (associated with the twisted field lines) under a dome-like magnetic field structure. (b) The filament starts to rise due to the loss of the force balance. Reconnection between highly sheared or twisted field below the filament occurs, resulting in the brightening under the filament. (c) When the filament is uplifted to a certain height, an impulsive reconnection in the null point is triggered and a circular flare ribbon appeared. During this process, a current sheet forms below the filament and magnetic reconnection occurred there may enhance the brightening below the filament. (d) As the filament is uplifted even higher, white-light emission enhancement appears at the two footpoints of the filament, possibly due to reconnection between the filament and the higher background magnetic field or reconnection within the filament. The brown polygons mark locations where magnetic reconnection might occur.}
\end{figure}

\end{document}